\documentclass[reprint,amsmath,amssymb,aps,prl,showpacs,twocolumn]{revtex4}
\usepackage[utf8]{inputenc}
\usepackage{graphicx}
\usepackage{dcolumn}
\usepackage{bm}
\usepackage[breaklinks]{hyperref}
\usepackage{threeparttable}
\usepackage{bbold}
\usepackage{subfigure}
\usepackage{color}
\usepackage[normalem]{ulem}
\usepackage{cancel}
\usepackage{braket}
\usepackage{mathrsfs}
\usepackage{mathtools}

\usepackage{booktabs}

\renewcommand*{\HyperDestNameFilter}[1]{\jobname-#1}

\newcommand\eqCRTA{\stackrel{\mathclap{\normalfont\mbox{CRTA}}}{=}}

\newcommand{\vn}[1]{{\boldsymbol{#1}}}

\begin{document}

\preprint{APS/123-QED}

\setcounter{secnumdepth}{2} 

\title{{Non-local field-like spin-orbit torques in Rashba systems:\\ an {\it ab-initio} study of Ag$_{2}$Bi/Ag/Fe film}}

\author{Guillaume G\'eranton}
\email{g.geranton@fz-juelich.de}
\author{Bernd Zimmermann}
\author{Nguyen H. Long}
\author{Phivos Mavropoulos}
\author{Stefan Bl\"ugel}
\author{Frank  Freimuth}
\email{f.freimuth@fz-juelich.de}
\author{Yuriy Mokrousov}
\email{y.mokrousov@fz-juelich.de}
\affiliation{Peter Gr\"unberg Institut and Institute for Advanced Simulation,
Forschungszentrum J\"ulich and JARA, 52425 J\"ulich, Germany}

\date{\today}

\begin{abstract}

We investigate from first principles the field-like spin-orbit torques (SOTs) in a Ag$_{2}$Bi-terminated Ag(111) film grown on ferromagnetic Fe(110). We find that a large part of the SOT arises from the spin-orbit interaction (SOI) in the Ag$_{2}$Bi layer far away from the Fe layers. These results clearly hint at a long range spin transfer in the direction perpendicular to the film that does not originate in the spin Hall effect. In order to bring evidence of the non-local character of the computed SOT, we show that the torque acting on the Fe layers can be engineered by the introduction of Bi vacancies in the Ag$_{2}$Bi layer. Overall, we find a drastic dependence of the SOT on the disorder type, which we explain by a complex interplay of different contributions to the SOT in the Brillouin zone.

\end{abstract}

\pacs{75.10.Lp, 03.65.Vf, 71.15.Mb, 71.20.Lp, 73.43.-f}
\maketitle

\section{Introduction}

Spin-orbit torques (SOTs) arise in non-centrosymmetric ferromagnetic thin films from the interplay of exchange and spin-orbit interactions~\cite{PhysRevB.79.094422,PhysRevB.80.134403,Miron:155001,Liu04052012,PhysRevLett.109.096602}. The origin of the SOT is usually attributed to two different mechanisms. The first one relies on the current-induced spin accumulation~\cite{V.M._Edelstein_1990,MihaiMiron:155006,Miron:154509,PhysRevB.85.180404,PhysRevB.87.174411} at the interface between the ferromagnet and the substrate, where magnetism, spin-orbit coupling (SOC) and broken inversion symmetry coexist. The second mechanism is attributed to the spin Hall effect~\cite{Kato10122004,RevModPhys.87.1213}, which generates a spin current injected into the ferromagnet. Among those two mechanisms, only the second one relies on a transfer of spin on a length scale larger than a few atomic distances and therefore has a non-local character.

In the past few years, the understanding of spin-orbit torques has been largely based on simple models. In this respect, the ferromagnetic Rasbha model~\cite{PhysRevB.79.094422} is very popular because it yields a simple electronic structure where the interplay between exchange and spin-orbit interactions can be well understood. However, the ferromagnetic Rashba model is not easily applicable to real systems, as the materials that exhibit clear Rashba-like bands are usually paramagnetic. Recently, a Ag$_{2}$Bi-terminated Ag(111) film grown on ferromagnetic Fe(110) has been proposed by Carbone {\it et al.}~\cite{PhysRevB.93.125409} as an ideal system to study the interplay of exchange and spin-orbit interactions. The strong $k$-asymmetry of the band structure was attributed to the spin-selective hybridization between Rashba-split surface states at the Ag$_{2}$Bi surface and exchange-split quantum well states in the Ag film. Apart from its fundamental interest for the study of systems lacking simultaneously time reversal and space inversion symmetry, the very peculiar electronic structure of that system suggests interesting properties in terms of spin-polarized transport. {This is strongly supported by the numerous experimental works on spin to charge conversion and spin pumping in similar systems that involve Ag/Bi interfaces~\cite{doi:10.1063/1.4919129,doi:10.1063/1.4921765,Nature_RS,doi:10.1063/1.4915479}.}

In this paper, we investigate from first principles the field-like SOT in a Ag$_{2}$Bi-terminated Ag(111) film grown on ferromagnetic Fe(110). In particular, we clarify the role of the Rashba states at the Ag$_{2}$Bi surface in giving rise to the torque and discuss the non-locality of the computed SOT. We stress the fact that this non-local torque does not originate in the intrinsic spin Hall effect, as it arises already in the semi-classical approach within the constant relaxation time approximation which does not account for the interband terms.

This paper is structured as follows. First, we show in Sec.~\ref{sec_elecstruct} the band structure of the film and discuss its symmetry. Next, we compute in Sec.~\ref{sec_CRTA} the components of the torkance tensor within the constant relaxation time approximation (CRTA) and discuss in detail the role of the Ag$_{2}$Bi layer in giving rise to the torque. Then, in Sec.~\ref{sec_disorder} we use our formalism for the extrinsic SOT to demonstrate the possibility to engineer the torque by tuning the disorder in the film. Finally, we conclude our study in Sec.~\ref{Conclusions}.

\section{Electronic structure}\label{sec_elecstruct}

\begin{figure}[]
\centering
\includegraphics*[width=8.5cm]{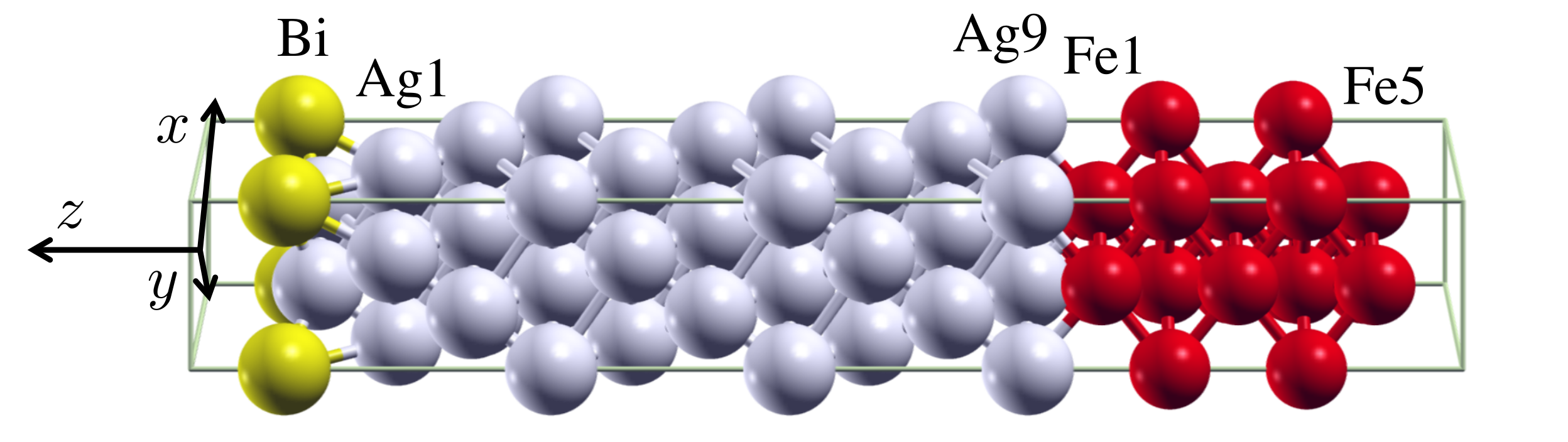}
\caption{Illustration of a unit cell of a Ag$_{2}$Bi-terminated Ag(111) film grown on ferromagnetic Fe(110). Bi, Ag and Fe atoms are represented in yellow, grey and red, respectively.}% Atomic coordinates for the 45 atoms in the unit cell are shown in Table~\ref{tab_atoms_BiAgFe}.}
\label{fig_illustration}
\end{figure}

We compute the electronic structure of the film using the Korringa-Kohn-Rostoker (KKR) method and the Vosko-Wilk-Nusair (VWN) functional. We have constructed a film that contains 5 layers of Fe(110), 9 layers of Ag(111) and one layer of  Ag$_{2}$Bi, which corresponds to the system studied in Ref.~\cite{PhysRevB.93.125409}. 
%The corresponding atomic coordinates are given in Table~\ref{tab_atoms_BiAgFe} along with the atomic magnetic moments. 
The magnetic moments of Fe atoms lie in the range of $2.69-2.95\,\mu_{B}$, while the induced magnetization in the Ag layers is negligible. The magnetization lies in the plane of the film and points in the $\vn{e}_{y}$ direction, see Fig.~\ref{fig_illustration}.

\begin{figure*}[]
\centering
\includegraphics*[width=6.55cm]{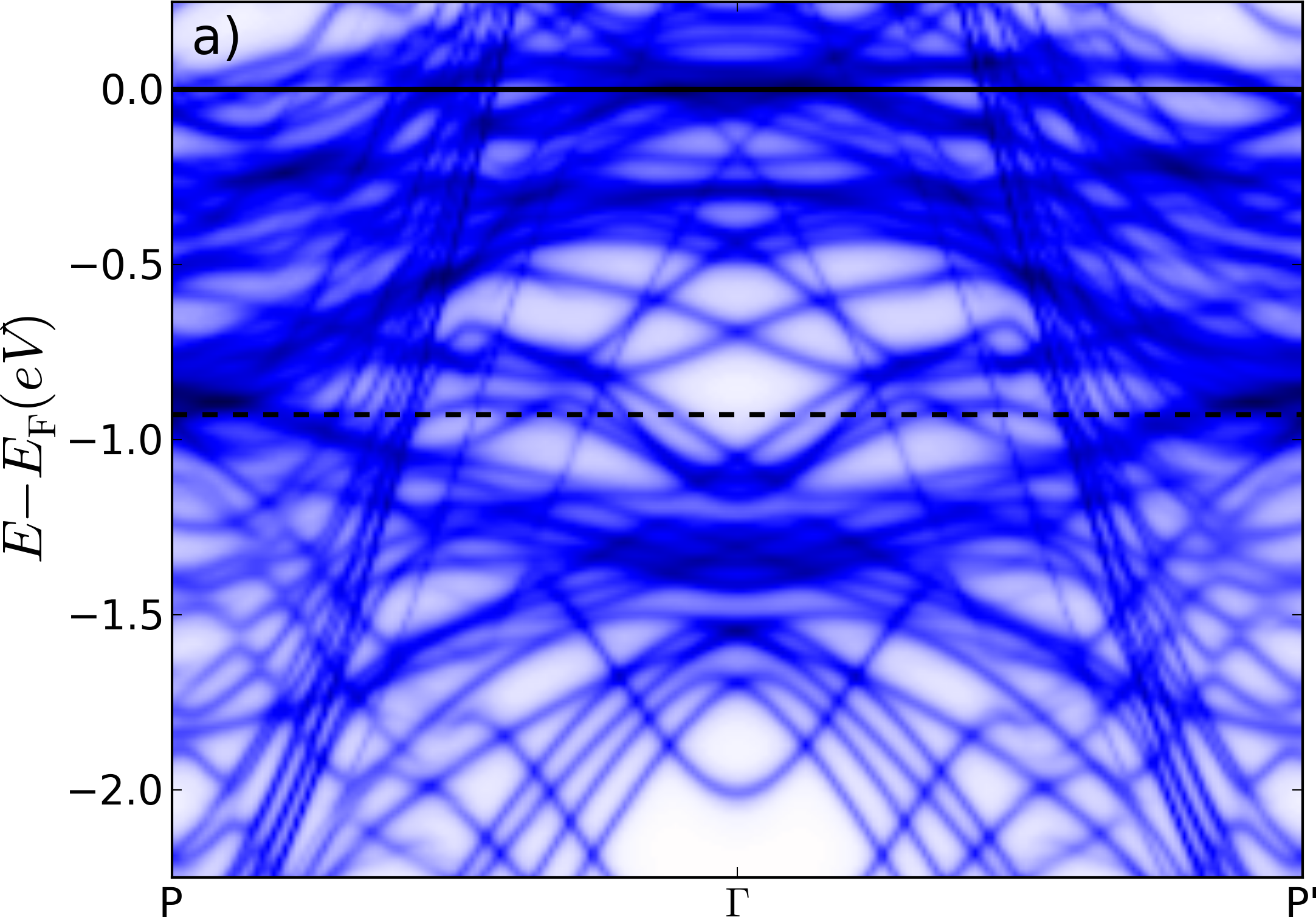}
\includegraphics*[width=6.55cm]{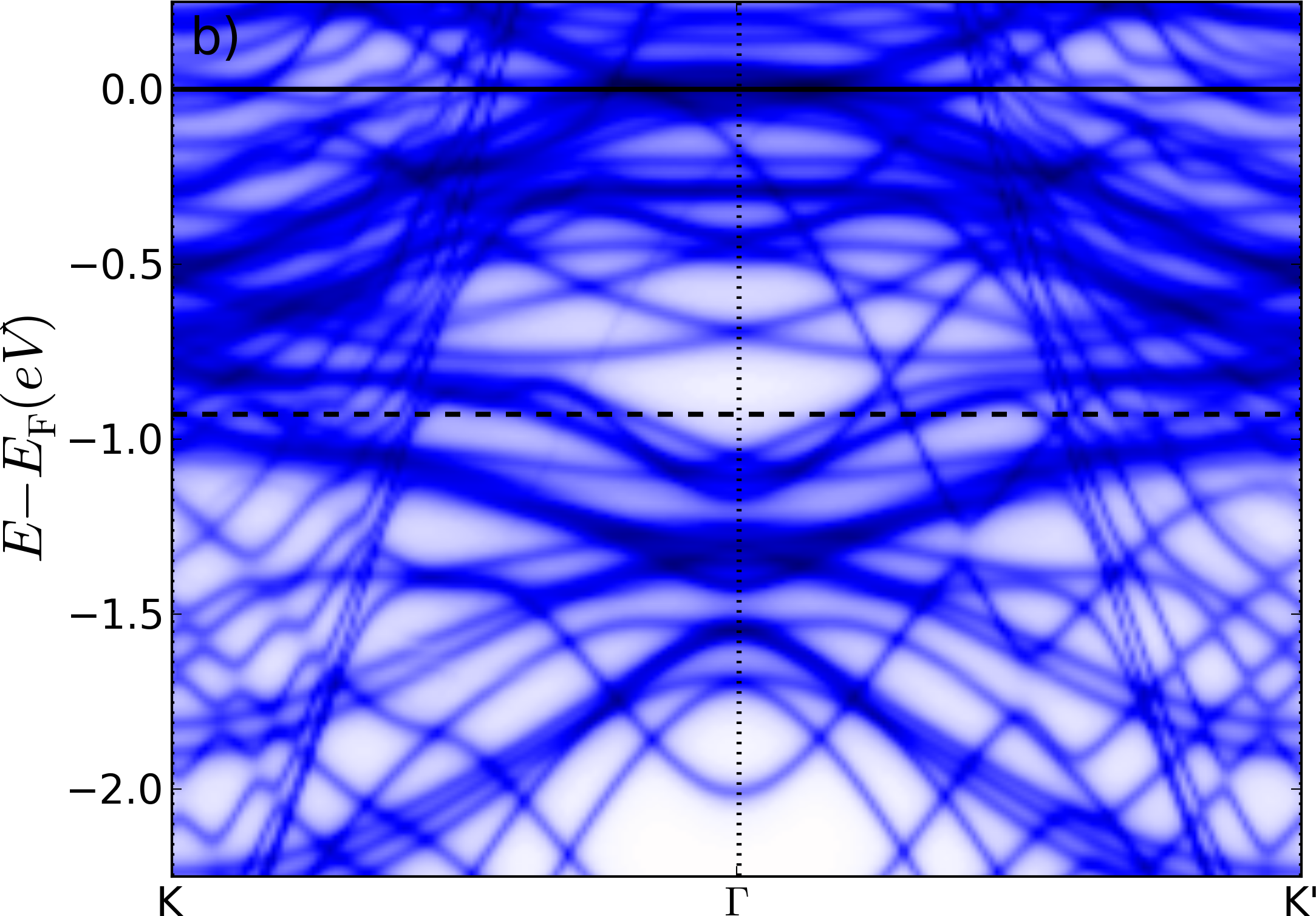}
\includegraphics*[width=6.55cm]{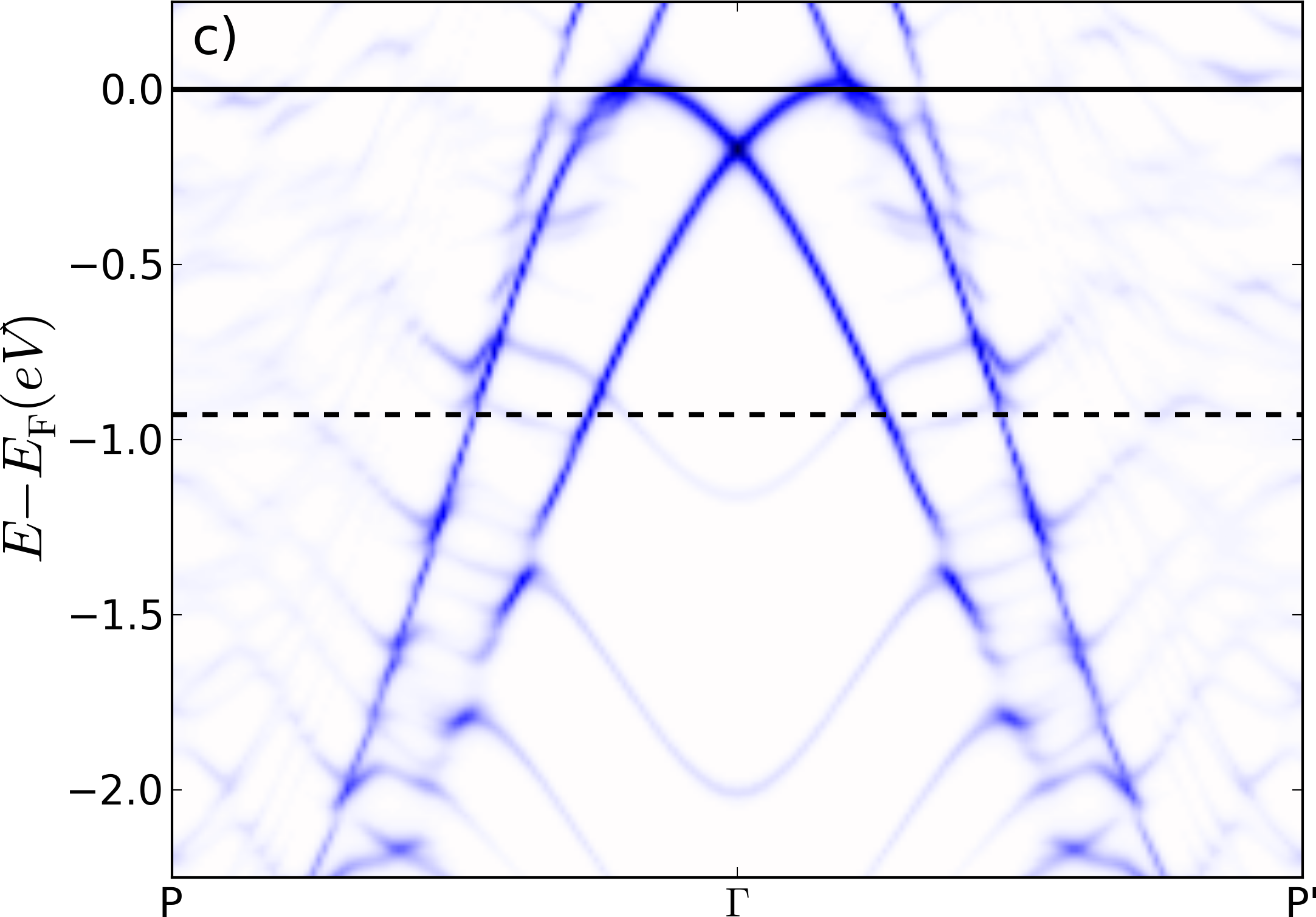}
\includegraphics*[width=6.55cm]{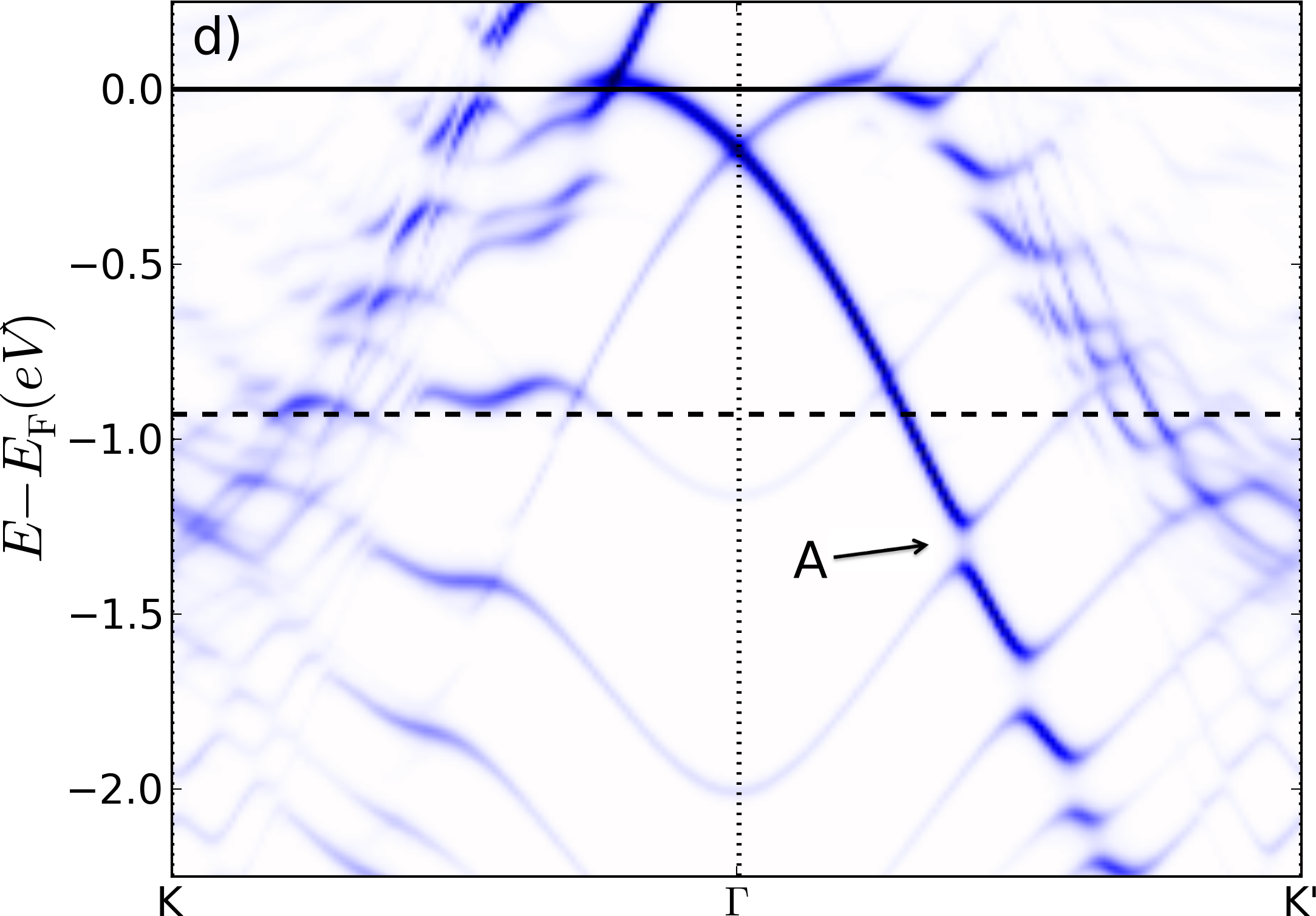}
\caption{(a,b) Spectral density of states for a Ag$_{2}$Bi-terminated Ag film grown on ferromagnetic Fe(110) along the $k$-paths P-$\Gamma$-P' and K-$\Gamma$-K' shown in Fig.~\ref{fig_BiAgFe_FS_we}. (c,d) Local density of states on the Ag$_{2}$Bi layer. An example of asymmetric band-gap opening is marked by the letter A. The dashed lines indicate the energy $E_{\rm S}=E_{\rm F}-0.93$\,eV corresponding to the states shown in Fig.~\ref{fig_BiAgFe_FS_we}b.}
\label{fig_BiAgFe_BS}
\end{figure*}

In Figs.~\ref{fig_BiAgFe_BS}a and~\ref{fig_BiAgFe_BS}b we display the $k$-resolved density of states (DOS) for $k$-vectors parallel and perpendicular to magnetization direction, respectively. In full argeement to Ref.~\cite{PhysRevB.93.125409}, we find that the band structure is symmetric for $k$-vectors parallel to magnetization direction (path P-$\Gamma$-P') and strongly asymmetric for $k$-vectors perpendicular to magnetization direction (path K-$\Gamma$-K'). 

We show in Figs.~\ref{fig_BiAgFe_BS}c and~\ref{fig_BiAgFe_BS}d the $k$-resolved local density of states (LDOS) for the Ag$_{2}$Bi layer. The Rashba bands are best visible along the path P-$\Gamma$-P' (Fig.~\ref{fig_BiAgFe_BS}c), where the energy splitting reaches as much as 0.8\,eV. The spin-selective hybridization between Rashba surface states and exchange-split quantum well states in the Ag film manifests itself most prominently in the opening of asymmetric band-gaps along the path K-$\Gamma$-K' (Fig.~\ref{fig_BiAgFe_BS}d). An example of asymmetric band-gap opening is marked by the letter A in Fig.~\ref{fig_BiAgFe_BS}d. From the $k$-resolved LDOS for the Ag$_{2}$Bi layer, it is clearly visible that the Fermi energy falls precisely at the top of the Rashba bands, which implies a relatively small total ($k$-integrated) density of states at the Fermi energy.

\begin{figure}[]
\centering
\includegraphics*[height=3.5cm]{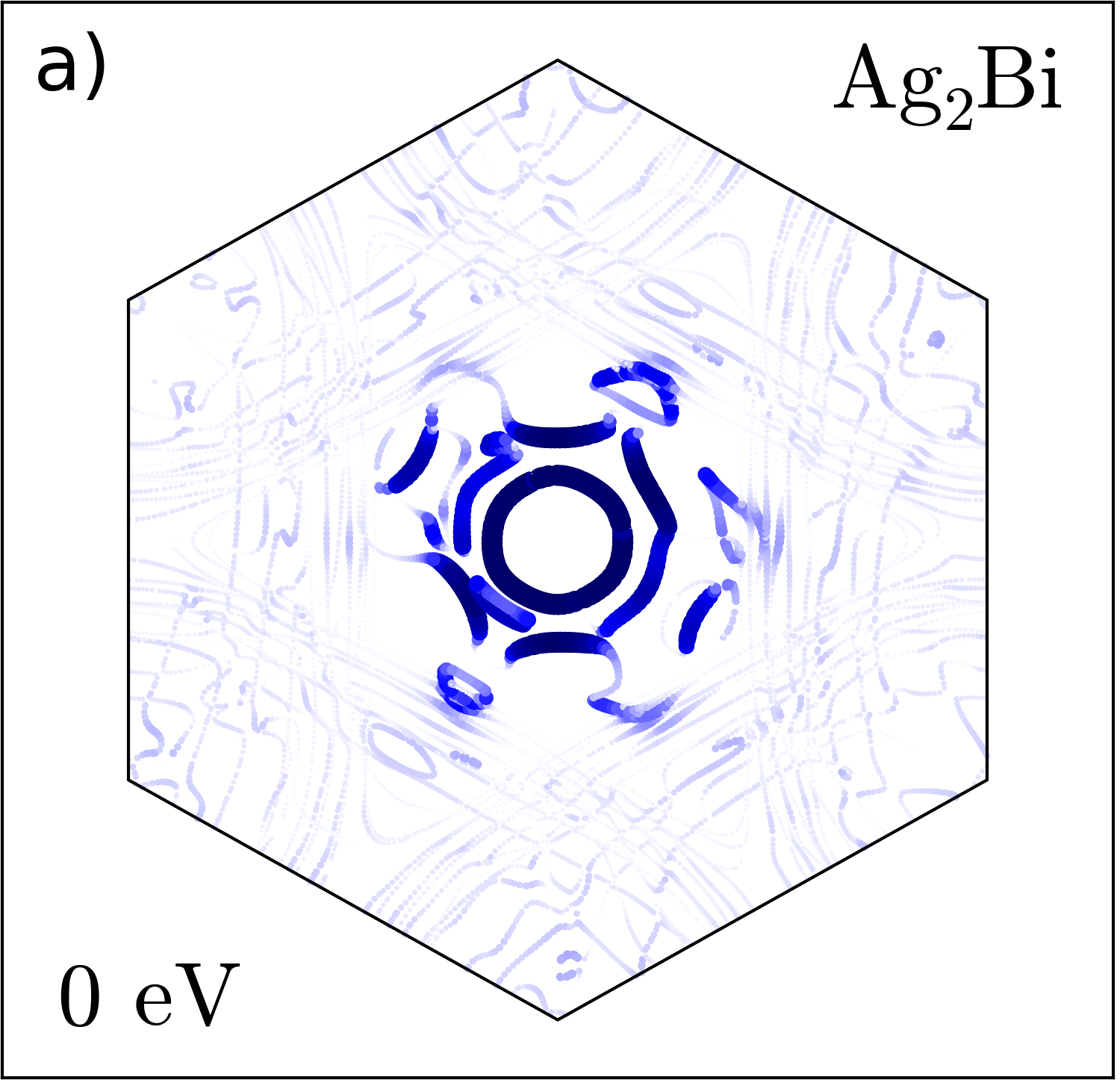}
\includegraphics*[height=3.5cm]{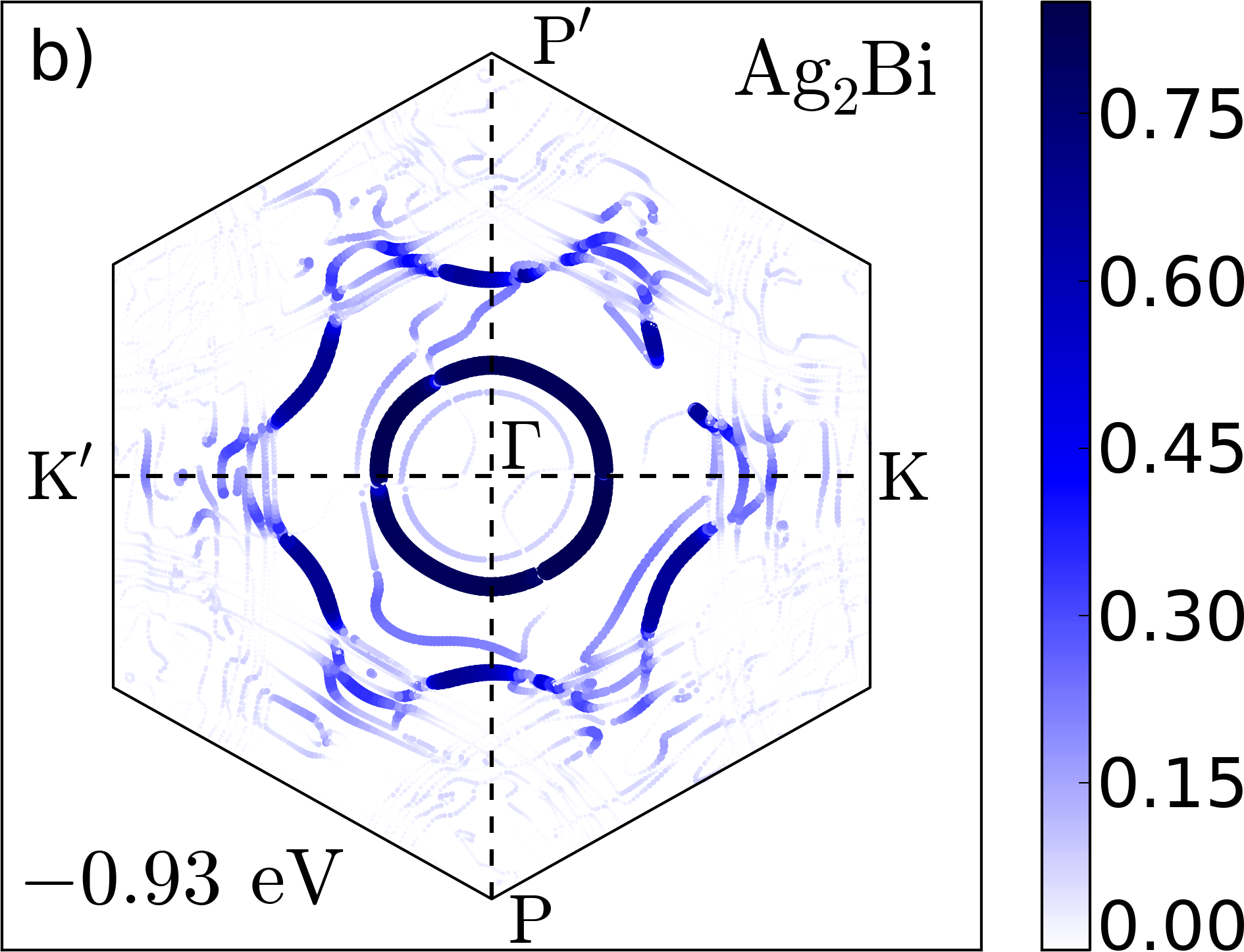}
\caption{(a) States at the Fermi energy $E_{\rm F}$ and (b) at the energy $E_{\rm S}=E_{\rm F}-0.93$\,eV below the Fermi energy marked by the portion of their wave function on the Ag$_{2}$Bi layer.}
\label{fig_BiAgFe_FS_we}
\end{figure}

We analyze the states at the Fermi energy in Fig.~\ref{fig_BiAgFe_FS_we}a, which we mark by their weight on the Ag$_{2}$Bi layer. The band closest to the $\Gamma$ point exhibits a nearly circular shape and the weight of the corresponding states  on the Ag$_{2}$Bi layer reaches as much as 75\%. The states found when moving away from the $\Gamma$ point are more strongly asymmetric and tend to have a smaller portion on the Ag$_{2}$Bi layer. Owing to the position of the Fermi energy with respect to the Rashba bands (see Fig.~\ref{fig_BiAgFe_BS}c), the states with a large weight on the Ag$_{2}$Bi layer cover a rather small part of the Brillouin zone (BZ).

Next, we analyze the states below the Fermi level. Namely, in Fig.~\ref{fig_BiAgFe_FS_we}b we display the states whose energy is lower by 0.93\,eV as compared to the Fermi energy and we mark them by the portion of their wave function on the Ag$_{2}$Bi layer. As compared to the case at the Fermi energy (Fig.~\ref{fig_BiAgFe_FS_we}a), we observe that the states with a larger weight on the Ag$_{2}$Bi layer spread over a much larger part of the BZ. This is in accordance with the dispersion of the Rashba bands in Figs.~\ref{fig_BiAgFe_BS}c and~\ref{fig_BiAgFe_BS}d. The strong asymmetry in $k$-space of the states with a large weight on the Ag$_{2}$Bi layer is a signature of the simultaneous breaking of space inversion symmetry and time reversal symmetry due to the strong exchange field in the Fe layers. The corresponding states can be expected to play an important role in giving rise to the torque on the magnetization, as their behaviour is strongly influenced by both spin-orbit and exchange interactions.

\section{Spin-orbit torques}
\label{sec_SOT}
In the presence of an external electric field $\vn{E}$, the crystal momentum of the electrons varies according to $\hbar \dot{\vn{k}}=-e\vn{E}$, where $e$ is the positive elementary charge. In order to predict experimental response functions such as longitudinal conductivities or torques, it is necessary to account for the collisions between electrons and, e.g., phonons or impurities. The vector mean free path of the electrons in the presence of disorder-induced electronic transitions can be obtained by solving the Boltzmann equation. The calculation of the transition rates $P_{\vn{k}\vn{k'}}$ and the vector mean free path $\vn{\lambda}(\vn{k})$ within the KKR method was discussed in Ref.~\cite{PhysRevLett.104.186403} and its application to the spin-orbit torques in Ref.~\cite{PhysRevB.93.224420}.

In this work, we use two different approximations for the calculation of the vector mean free path. First, we neglect the scattering-in term in the Boltzmann equation, which is important for the calculation of the spin or anomalous Hall angles, but not for the longitudinal conductivities or the extrinsic spin-orbit torques. This yields an explicit expression for the vector mean free path
\begin{equation}\label{BOLTZ}
\mbox{{\boldmath$\lambda$}}(\vn{k}) = \tau_{\vn{k}}\vn{v}(\vn{k}),
\end{equation}
where $\vn{v}(\vn{k})$ is the group velocity and $\tau_{\vn{k}}$ is the state-dependent relaxation time obtained as
\begin{equation}\label{relax_time}
\tau_{\vn{k}}^{-1}=\sum_{\vn{k'}}P_{\vn{k'}\vn{k}}.
\end{equation}
On the one hand, this approximation has the great advantage that it does not require to solve iteratively the Boltzmann equation, which causes convergence problems for systems lacking symmetry. On the other hand, it reproduces very accurately the dependence of the SOT on the type of disorder, which arises essentially from the modulation of the relaxation times for different states in the BZ.

The second level of approximation is the constant relaxation time approximation (CRTA), where $\tau_{\vn{k}}$ is set to a constant value $\tau^{0}$, i.e.,
\begin{equation}\label{const_relax_time}
\tau_{\vn{k}}~~~ \eqCRTA ~~~\tau^{0} = \frac{\hbar}{2 \Gamma}.
\end{equation}
The energy parameter $\Gamma$ quantifies the disorder strength and we set it to be 25\,meV in order to mimic the effect of room temperature. The advantage of the CRTA is that the dependence of the spin-orbit torques on the disorder strength is known analytically and it does not require the calculation of the transition rates $P_{\vn{k}\vn{k'}}$. The disadvantage is that it completely looses the dependence of the SOT on the type of disorder.

{From the knowledge of the vector mean free path of the electrons $\vn{\lambda}(\vn{k})$, we compute the spin accumulation~$\vn{s}$ and the spin-orbit torque~$\vn{T}$ induced by an external electric field $\vn{E}$. We define the response tensors for the spin accumulation, $\vn{\chi}$, and the torkance, $\vn{t}$, according to $\vn{s} = \bm{\chi}\vn{E}$ and $\vn{T} = \vn{t}\vn{E}$.
%\begin{equation}\label{BOLTZ_safin3_3D}
%\vn{s} = \bm{\chi}\vn{E}
%\end{equation}
%\begin{equation}\label{BOLTZ_Tfin3_3D}
%\vn{T} = \vn{t}\vn{E}.
%\end{equation}
The two response tensors are given by
\begin{equation}\label{BOLTZ_sa_2D}
\bm{\chi} = \frac{e \mu_{B}}{\hbar \mathcal{S_{BZ}}} \int_{\rm FS} \frac{dk}{|\vn{v}(\vn{k})|}~\langle \bm{\sigma} \rangle_{\vn{k}} \otimes \mbox{{\boldmath$\lambda$}}(\vn{k})
\end{equation}
and
\begin{equation}\label{BOLTZ_t_2D_atom}
\vn{t} = \frac{e}{\hbar \mathcal{S_{BZ}}} \int_{\rm FS} \frac{dk}{|\vn{v}(\vn{k})|}~\langle \bm{\mathcal{T}} \rangle_{\vn{k}} \otimes  \mbox{{\boldmath$\lambda$}}(\vn{k}),
\end{equation}
where $\mathcal{S_{BZ}}$ is the area of the two-dimensional BZ. The torque operator $\bm{\mathcal{T}}(\vn{r})$ is given by the vector product of the spin magnetic moment $-\mu_{B}\boldsymbol{\sigma}$ with the exchange field $\vn{B}^{\rm xc}(\vn{r})$, where $\boldsymbol{\sigma}$ is the vector of Pauli matrices. The calculation of the matrix elements for the torque operator within the KKR method was discussed in Ref.~\cite{PhysRevB.93.224420}.

In the following, we first investigate in Sec.~\ref{sec_CRTA} the SOTs in a Ag$_{2}$Bi/Ag/Fe thin film within the CRTA, i.e., using Eqs.~\ref{BOLTZ}, \ref{const_relax_time} and~\ref{BOLTZ_t_2D_atom}, in order to clarify the role of the Ag$_{2}$Bi layer in giving rise to the torque on the magnetization. Then, we compute in Sec.~\ref{sec_disorder} the SOTs based on the first principles transition rates $P_{\vn{k'}\vn{k}}$ in the presence of Bi vacancies in the Ag$_{2}$Bi layer using Eqs.~\ref{BOLTZ}, \ref{relax_time} and~\ref{BOLTZ_t_2D_atom}.}

\subsection{Spin-orbit torques within the constant relaxation time approximation}
\label{sec_CRTA}

\subsubsection{Torkance at the true Fermi energy $E_{\rm F}$}\label{sec_torkEf}
\begin{figure}[]
\centering
\includegraphics*[height=3.5cm]{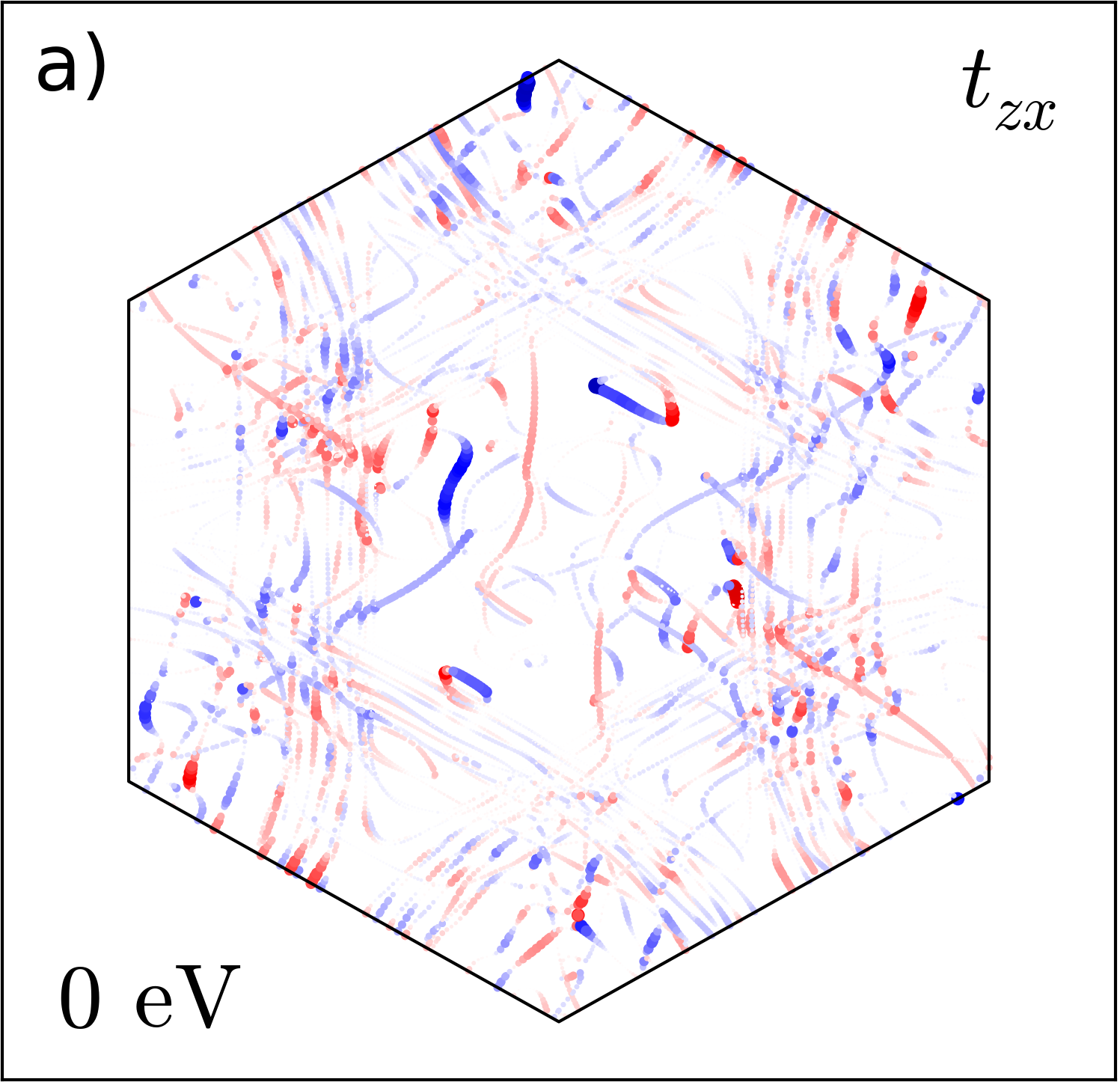}
\includegraphics*[height=3.5cm]{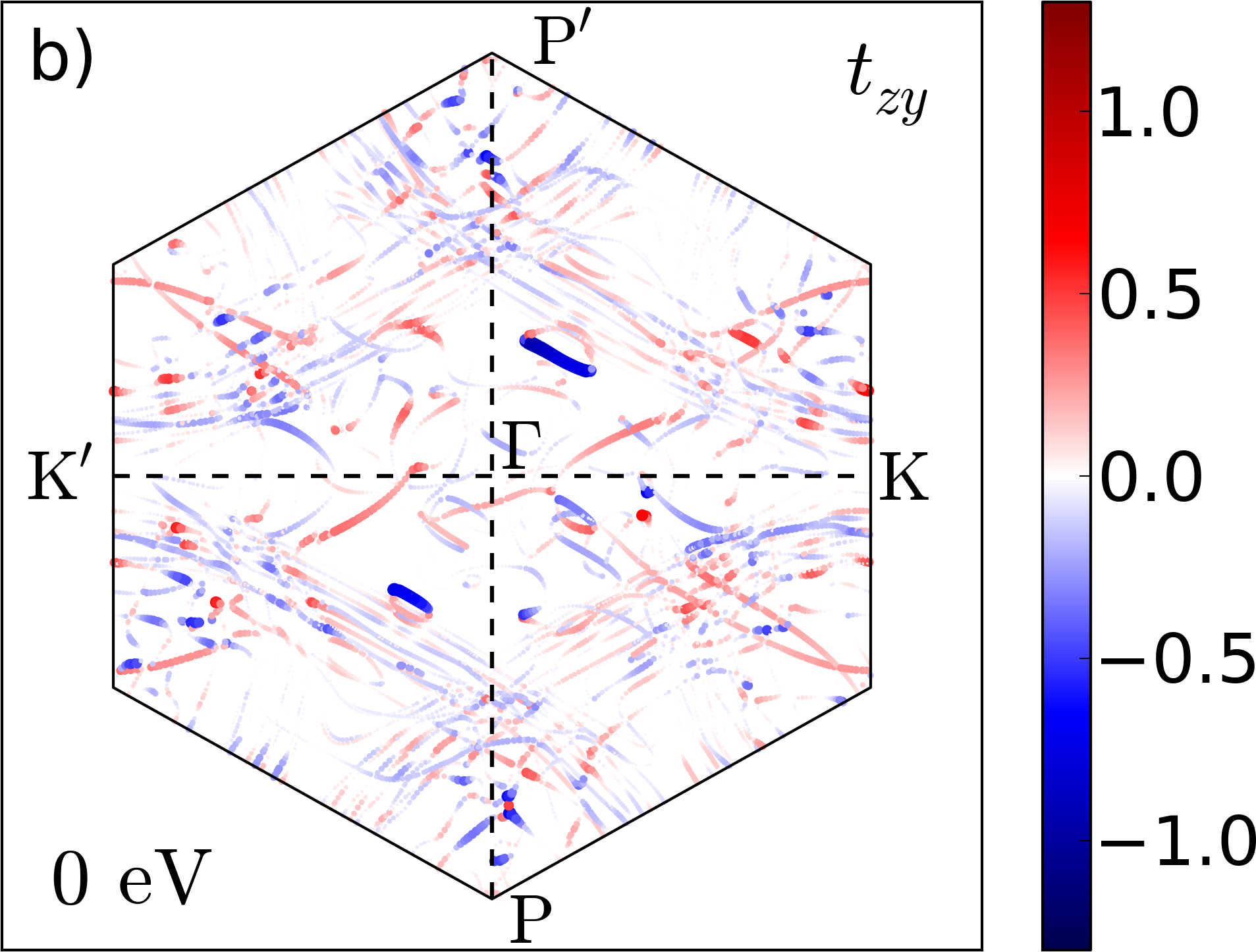}
\caption[Torkance at the Fermi surface of a Fe/Ag/Ag$_{2}$Bi film]{Distribution of $\mathcal{T}_{z}(\vn{k}) \lambda_{x}(\vn{k})/|\vn{v}(\vn{k})|$ and $\mathcal{T}_{z}(\vn{k}) \lambda_{y}(\vn{k})/|\vn{v}(\vn{k})|$ in the Brillouin zone for the Bi/Ag/Fe film. The Fermi energy was set to its true value $E_{\rm F}$ {and the disorder strength to $\Gamma = 25$\,meV.}}
\label{fig_BiAgFe_t_Ef}
\end{figure}

We first compute the torkance of the Ag$_{2}$Bi/Ag/Fe thin film with the Fermi energy set to its true value $E_{\rm F}$. We use the constant relaxation time approximation where the vector mean free path is given by $\vn{\lambda}(\vn{k})=(2\Gamma /\hbar)\, \vn{v}(\vn{k})$ with $\Gamma=25$\,meV. We find two non-vanishing components of the torkance tensor $t_{zx}=-0.18$\,ea$_0$ and $t_{zy}=-0.37$\,ea$_0$. As shown in Figs.~\ref{fig_BiAgFe_BS}c, \ref{fig_BiAgFe_BS}d and~\ref{fig_BiAgFe_FS_we}, the density of Rashba states at the true Fermi energy is relatively small, so that their role in mediating the computed torkance in this case is expected to be minor. 
Indeed, as we show in Fig.~\ref{fig_BiAgFe_t_Ef}, for both $t_{zx}$ and~$t_{zy}$ components the values of the torkance are a results of a complex competition of the bands spread throughout the Brillouin zone. Comparing the distribution of the torkance with the weights of the states on the Ag$_{2}$Bi layer (Fig.~\ref{fig_BiAgFe_FS_we}a), we observe no correlation between both quantities. Only very small pockets of states with large weights on the Ag$_{2}$Bi layer yield a sizeable contribution to the torkance. Overall, the distribution of the torkance in the Brillouin zone shows that the SOI in the Ag$_{2}$Bi layer is not the main mechanism giving rise to the SOT when the Fermi energy is set to its true value. However, the dispersion of the Rashba bands (Figs.~\ref{fig_BiAgFe_BS}c and~\ref{fig_BiAgFe_BS}d) suggests that a much stronger effect of the corresponding states could be expected when the Fermi energy is shifted below its true value. {This could be realized, for example, by gating the system, tuning the  thickness of the Ag layer, or alloying  Bi layer with other elements with strong spin-orbit strength, such as~e.g.~Pb.}

\subsubsection{Torkance at the Fermi energy set to $E_{\rm S}=E_{\rm F}-0.93$\,eV}\label{sec_torkEf_shifted}
\begin{figure}[]
\centering
\includegraphics*[height=3.5cm]{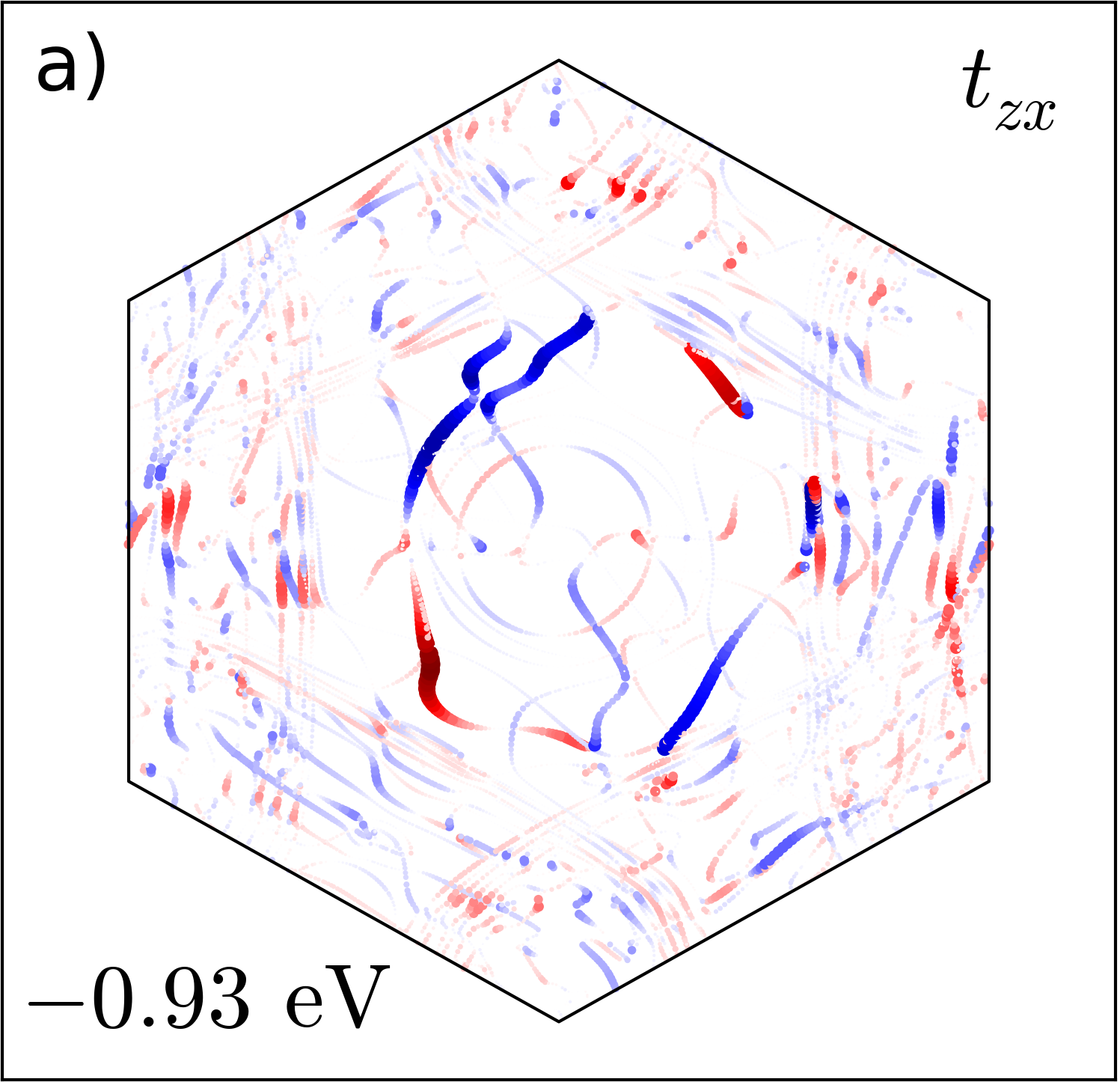}
\includegraphics*[height=3.5cm]{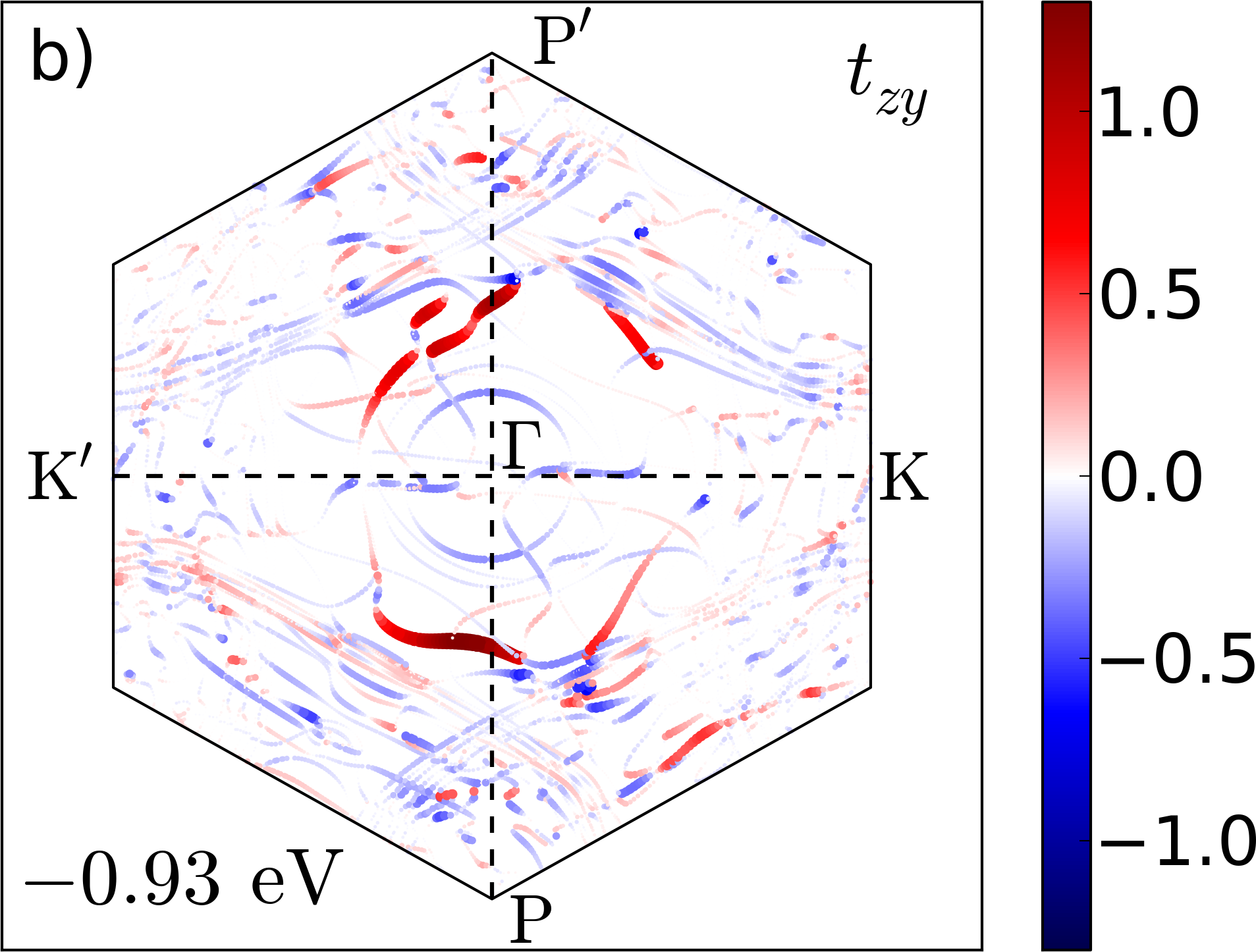}
\caption[Torkance 930\,meV below the Fermi energy in a Fe/Ag/Ag$_{2}$Bi film]{Distribution of $\mathcal{T}_{z}(\vn{k}) \lambda_{x}(\vn{k})/|\vn{v}(\vn{k})|$ and $\mathcal{T}_{z}(\vn{k}) \lambda_{y}(\vn{k})/|\vn{v}(\vn{k})|$ in the Brillouin zone for the Bi/Ag/Fe film. The Fermi energy was set to $E_{\rm S}=E_{\rm F}-0.93$\,eV {and the disorder strength to $\Gamma = 25$\,meV.}}
\label{fig_BiAgFe_t_Ef_min}
\end{figure}

In case when the Fermi energy is set to $E_{\rm S}=E_{\rm F}-0.93$\,eV,  as apparent from Fig.~\ref{fig_BiAgFe_t_Ef_min}, the distribution of the torkance in the Brillouin zone is also very complicated. However, a striking difference to the previous case is the presence of a collection of states around the $\Gamma$-point with considerably larger contribution to the torkance. Comparing Fig.~\ref{fig_BiAgFe_t_Ef_min} and Fig.~\ref{fig_BiAgFe_FS_we}b, it becomes immediately clear that this collection of states hybridizes to some extent with the Rasbha states, having a weight of about 15\% on the Ag$_{2}$Bi layer. This observation is a first indication of the crucial role of the SOI in the Ag$_{2}$Bi layer in giving rise to the SOT in our system.

A second evidence of the role of the SOI in the Ag$_{2}$Bi layer in giving rise to the SOT can be obtained by comparing the spin accumulation profile and the spin-orbit torques to the case where the Ag$_{2}$Bi layer is removed. To do this, we have performed a similar DFT calculation for a Ag/Fe thin film constructed from 9 layers of Ag(111) and 5 layers of Fe(110). The layer-resolved response coefficients for the spin accumulation and the torque are given in Table~\ref{tab_intval_noBiAg}. We find for the torkance in the Fe layers a very large component $t_{zy}=-1.71$\,ea$_0$ and a rather small component $t_{zx}=-0.09$\,ea$_0$, which corresponds to the values of the effective magnetic field $B_x$ which acts on the magnetization of the order of
 0.57 and 0.02\,mT for a current density of $10^{7}$\,A/cm$^{2}$, respectively. As for the spin accumulation, it is of rather moderate magnitude in all regions of the film.

\begin{table}[t!]
\begin{tabular}{cccccc}\toprule
                        &$\chi_{xy}$& $\chi_{xx}$ & $t_{zy}$ & $t_{zx}$\\
\hline
Ag1-Ag4                     & \phantom{$+$}1.82 & \phantom{$+$}0.93 & \phantom{$+$}0.00  & \phantom{$+$}0.00\\
Ag5-Ag9   & \phantom{$+$}1.14 & \phantom{$+$}1.12 &  \phantom{$+$}0.00 & \phantom{$+$}0.01\\ 
Fe1-Fe5   & \phantom{$+$}1.32  & \phantom{$+$}0.10 & {\bf $-$1.71}                    & $-$0.09\\\bottomrule
\end{tabular}
\caption{
Values of the response coefficients for the spin accumulation (in units of $10^{-8}\,\mu_{\rm B}$/(V/cm)) and the torque (in units of ea$_0$) over three regions of a Ag/Fe thin film. The Fermi energy is set to $E_{\rm S}=E_{\rm F}-0.93$\,eV. Values are given for a super cell of three atoms per atomic layer in order to make easier the comparison with the Ag$_{2}$Bi/Ag/Fe thin film (see Table.~\ref{tab_intval})
}
\label{tab_intval_noBiAg}
\end{table}
The addition of the Ag$_{2}$Bi layer on top of the Ag layers drastically changes the spin accumulation profile in the film, as shown in Table~\ref{tab_intval}. The component $\chi_{xy}$ in the Ag$_{2}$Bi-Ag5 region changes sign and its magnitude increases by a factor of 6 as compared to the case of the Ag1-Ag4 region of the Ag/Fe film. The diffusion of the spins towards the bottom of the film strongly influences the spin accumulation in the Ag5-Ag9 and Fe1-Fe5 regions, where the $\chi_{xy}$ component equals $-2.10$ and $0.65\times 10^{-8}\mu_{\rm B}$(V/cm), respectively. As for the $\chi_{xx}$ component, it increases by about 20\% as compared to the case of the Ag1-Ag4 region of the Ag/Fe film. This is consistent with the symmetry of the spin accumulation expected from the paramagnetic Rashba model, i.e., $\chi_{xx} << \chi_{xy}$. Despite a large separation between the Ag$_{2}$Bi layer and Fe layers, the torque exerted on the ferromagnet is strongly influenced by the diffusion of the spins from the Ag$_{2}$Bi layer towards the bottom of the film. This results in the torkance components $t_{zx}=-0.32$\,ea$_0$ and $t_{zy}=-0.66$\,ea$_0$ for the Fe layers being respectively by a factor of 3.6 larger and by a factor of 2.6 smaller, as compared to the case of the Ag/Fe film. Therefore, both the magnitude and symmetry of the torkance tensor are strongly influenced by the presence of the Ag$_{2}$Bi layer. The crucial role of the collection of states that hybridize with the Rashba states in the Ag$_{2}$Bi layer suggests that a state-dependent scattering mechanism might be used to modulate the torkance tensor by either promoting or suppressing the contribution of the Rashba states  to the SOT.

\begin{table}[t!]
\begin{tabular}{cccccc}\toprule
                        &$\chi_{xy}$& $\chi_{xx}$ & $t_{zy}$ & $t_{zx}$\\
\hline
Ag$_{2}$Bi-Ag4          &{\bf $-$11.3}& \phantom{$+$}1.13                & \phantom{$+$}0.00   & \phantom{$+$}0.00\\
\phantom{g$_{2}$}Ag5-Ag9   & $-$2.10 & \phantom{$+$}1.00                      & $-$0.01   & \phantom{$+$}0.00\\
\phantom{g$_{2}$}Fe1-Fe5                      & \phantom{$+$}0.65  & \phantom{$+$}0.29  & {\bf $-$0.66} & {\bf $-$0.32}\\\bottomrule
\end{tabular}
\caption{
Values of the response coefficients for the spin accumulation (in units of $10^{-8}\,\mu_{\rm B}$/(V/cm)) and the torque (in units of ea$_0$) over three regions of a Ag$_{2}$Bi/Ag/Fe thin film. The Fermi energy is set to $E_{\rm S}=E_{\rm F}-0.93$\,eV.
}
\label{tab_intval}
\end{table}

\subsection{SOTs based on the first principles transition rates in the presence of Bi vacancies}\label{sec_disorder}
In this section we investigate the SOT in a Ag$_{2}$Bi/Ag/Fe thin film based on the first principles transition rates induced by Bi vacancies in the Ag$_{2}$Bi layer. In order to account for other sources of scattering that are present at finite temperature, such as phonons, we add to the first principles transition rates $P_{\vn{k'}\vn{k}}^{\rm imp}$ a constant diagonal contribution characterized by the disorder strength $\Gamma$. According to Ref.~\cite{PhysRevB.93.224420}, the total transition rates in that case are given by 
\begin{equation}\label{compPkk}
P_{\vn{k'}\vn{k}}=P_{\vn{k'}\vn{k}}^{\rm imp} + \frac{2\Gamma}{\hbar n(\varepsilon(\vn{k'}))} \delta(\varepsilon(\vn{k})-\varepsilon(\vn{k'})),
\end{equation}
where $n(\varepsilon(\vn{k}))$ is the density of states. The corresponding relaxation times are defined by
\begin{equation}\label{comptauk}
\frac{1}{\tau_{\vn{k}}} = \frac{1}{\tau^{\rm imp}_{\vn{k}}} + \frac{2\Gamma}{\hbar}\end{equation}
with $(\tau^{\rm imp}_{\vn{k}})^{-1}=\sum_{\vn{k'}}P_{\vn{k'}\vn{k}}^{\rm imp}$. In the case of the CRTA, i.e., if the impurity concentration $\bar{c}_{imp}=0$, the torkance is inversely proportional to the disorder strength $\Gamma$. In the case when the impurity-induced transition rates $P_{\vn{k'}\vn{k}}^{\rm imp}$ are the only source of scattering, i.e., if $\Gamma = 0$, the torkance is inversely proportional to the impurity concentration $\bar{c}_{imp}$. However, in the general case, the torkance tensor has a non-trivial dependence on both $\Gamma$ and $\bar{c}_{\rm imp}$.

\begin{figure}[t!]
\centering
\includegraphics*[height=4.3cm]{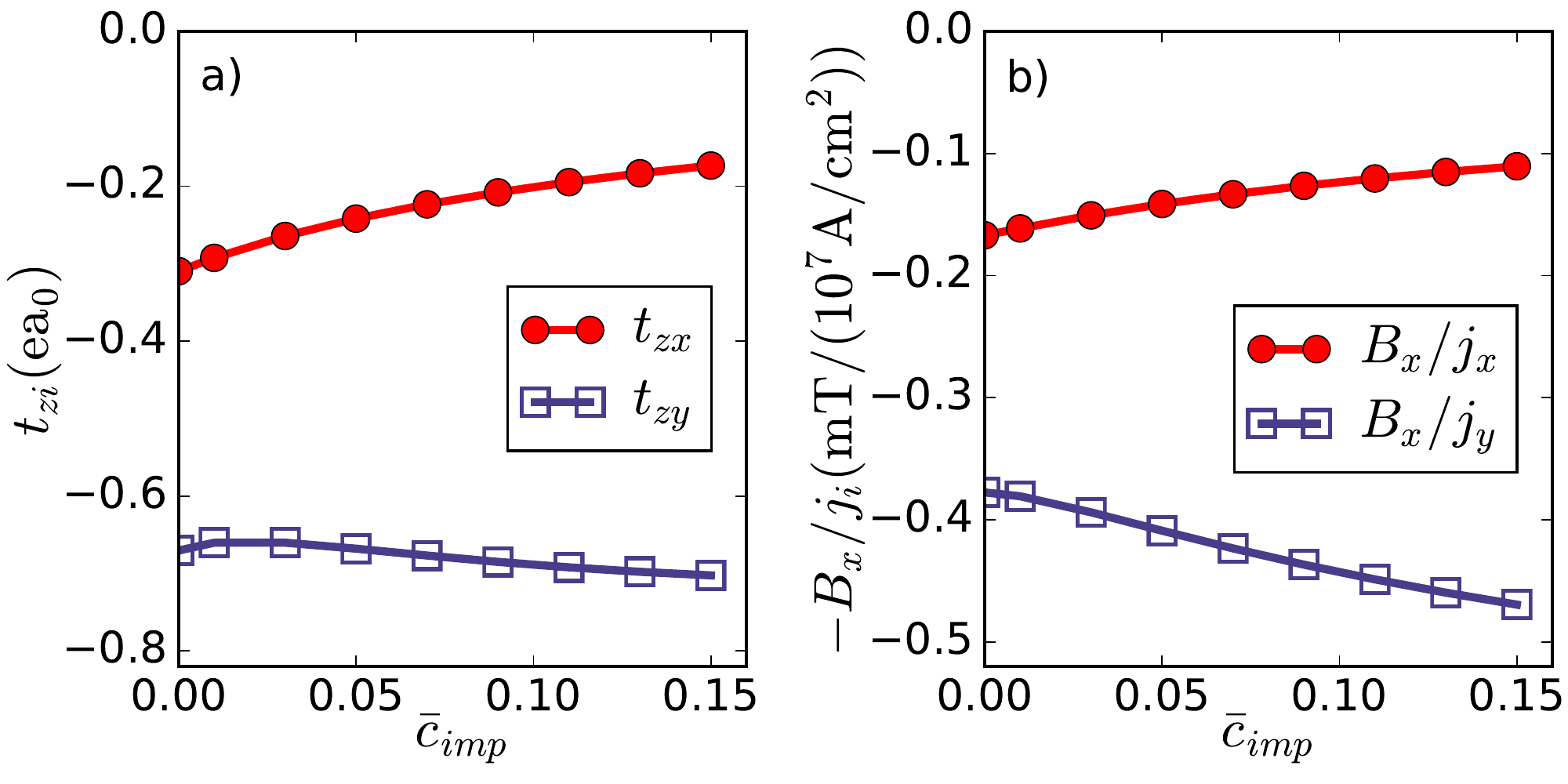}
\caption{(a) Torkance and (b) effective magnetic fields per unit of current density as a function of the impurity concentration $\bar{c}_{imp}$ for $\Gamma = 25$\,meV.}
\label{fig_BiAgFe_Beff_vs_nimp}
\end{figure}

We show in Fig.~\ref{fig_BiAgFe_Beff_vs_nimp}a the torkance components $t_{zx}$ and $t_{zy}$ as a function of impurity concentration for a given value of disorder strength $\Gamma = 25$\,meV. The torkance component $t_{zx}$ decreases continuously from its CRTA value ($\bar{c}_{imp}=0$) within the range of impurity concentration of 0$-$0.15. This trend can be explained by the reduction of the average relaxation time, and it is also expected for the longitudinal conductivity. However, the torkance component $t_{zy}$ decreases in the range of impurity concentration of 0$-$0.01 but increases in the range of 0.01$-$0.15. This behaviour can not be explained solely by the reduction of the average relaxation time, which would necessarily tend to decrease the torkance. Instead, it is necessary to consider the dispersion of the relaxation times $\tau_{\vn{k}}$ over the BZ, which tends to suppress the contributions of the states with large weights on the Ag$_{2}$Bi layer. The suppression of the large positive contribution from these states (Fig.~\ref{fig_BiAgFe_t_Ef_min}) results in a net increase of the total (negative) $t_{zy}$ component of the torkance tensor in the range of 0.01$-$0.15, as observed in Fig.~\ref{fig_BiAgFe_Beff_vs_nimp}a.

In Fig.~\ref{fig_BiAgFe_Beff_vs_nimp}b we present the values the effective magnetic fields per unit of current density for varying impurity concentration and a constant disorder strength $\Gamma = 25$\,meV. The effective magnetic fields follow the general trend of the torkance, see Fig.~\ref{fig_BiAgFe_Beff_vs_nimp}a. For a current density $j_{x}=10^{7}$\,A/cm$^{2}$  along the $x$-direction, the effective field $B_{x}$ varies from $-$0.17 to $-$0.11 mT over the range of impurity concentration of 0$-$0.15. For a current density $j_{y}$ of the same magnitude but applied along the $y$-direction, the effective field $B_{x}$ varies from $-$0.38 to $-$0.47 mT over the same range of impurity concentration. Therefore, increasing the concentration of Bi vacancies by 0.01 yields a variation of the effective magnetic fields of about 2\%, which is remarkable given the large distance between the Ag$_{2}$Bi and Fe layers.

\begin{figure}[t!]
\centering
\includegraphics*[height=4.3cm]{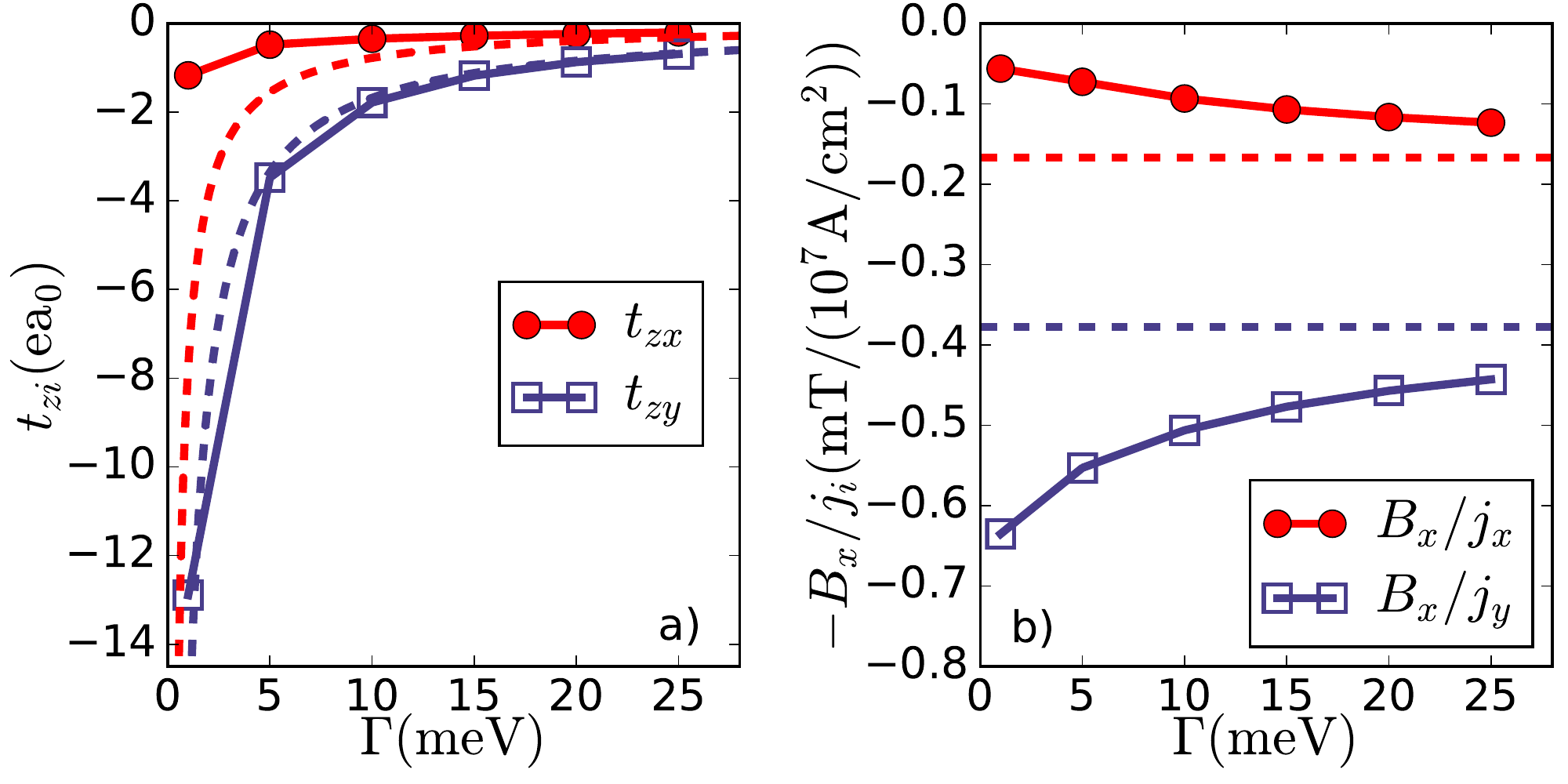}
\caption{(a) Torkance and (b) effective magnetic fields per unit of current density as a function of the disorder strength $\Gamma$ for $\bar{c}_{imp}=0$ (dashed lines) and $\bar{c}_{imp}=0.1$ (full lines).}
\label{fig_BiAgFe_Beff_vs_gamma}
\end{figure}

We compare in Fig.~\ref{fig_BiAgFe_Beff_vs_gamma}a the torkance components in cases $\bar{c}_{imp}=0$ (CRTA) and $\bar{c}_{\rm imp}=0.1$ for varying disorder strength $\Gamma$. In the range of disorder strength of 5$-$25\,meV, the torkance follows the general trend observed in Fig.~\ref{fig_BiAgFe_Beff_vs_nimp}a, i.e., the $t_{zx}$ ($t_{zy}$) component decreases (increases) in the presence of a finite concentration of Bi vacancies as compared to the CRTA case. For $\Gamma=1$\,meV both components are smaller than their corresponding CRTA values, which is consistent with the divergence expected for $\Gamma \rightarrow 0$ in the CRTA case.

The effective magnetic fields per unit of current density in the case when $\bar{c}_{imp}=0$ (CRTA) and $\bar{c}_{\rm imp}=0.1$ for varying disorder strength $\Gamma$ are shown in Fig.~\ref{fig_BiAgFe_Beff_vs_gamma}b. In the CRTA case, the ratio of the effective field to the current density is independent of the disorder strength, since both quantities are inversely proportional to $\Gamma$. In the case when $\bar{c}_{\rm imp}=0.1$, the effective field follows the general trend observed in Fig.~\ref{fig_BiAgFe_Beff_vs_nimp}b, i.e., it  decreases (increases) as compared to the CRTA case when the current density $\vn{j}$ is applied along the $x$($y$)-directions. The deviation from the CRTA values is most pronounced for low values of disorder strength. At $\Gamma=1$\,meV, the effective magnetic field is respectively reduced or increased by about 70\% as compared to its CRTA value, when the current density is injected in the $x$- or $y$-direction. Around $\Gamma=25$\,meV, the effect of the Bi vacancies is still very sizeable and can be tuned by varying the impurity concentration, see Fig.~\ref{fig_BiAgFe_Beff_vs_nimp}b. 

\begin{figure}[t!]
\centering
\includegraphics*[height=3.5cm]{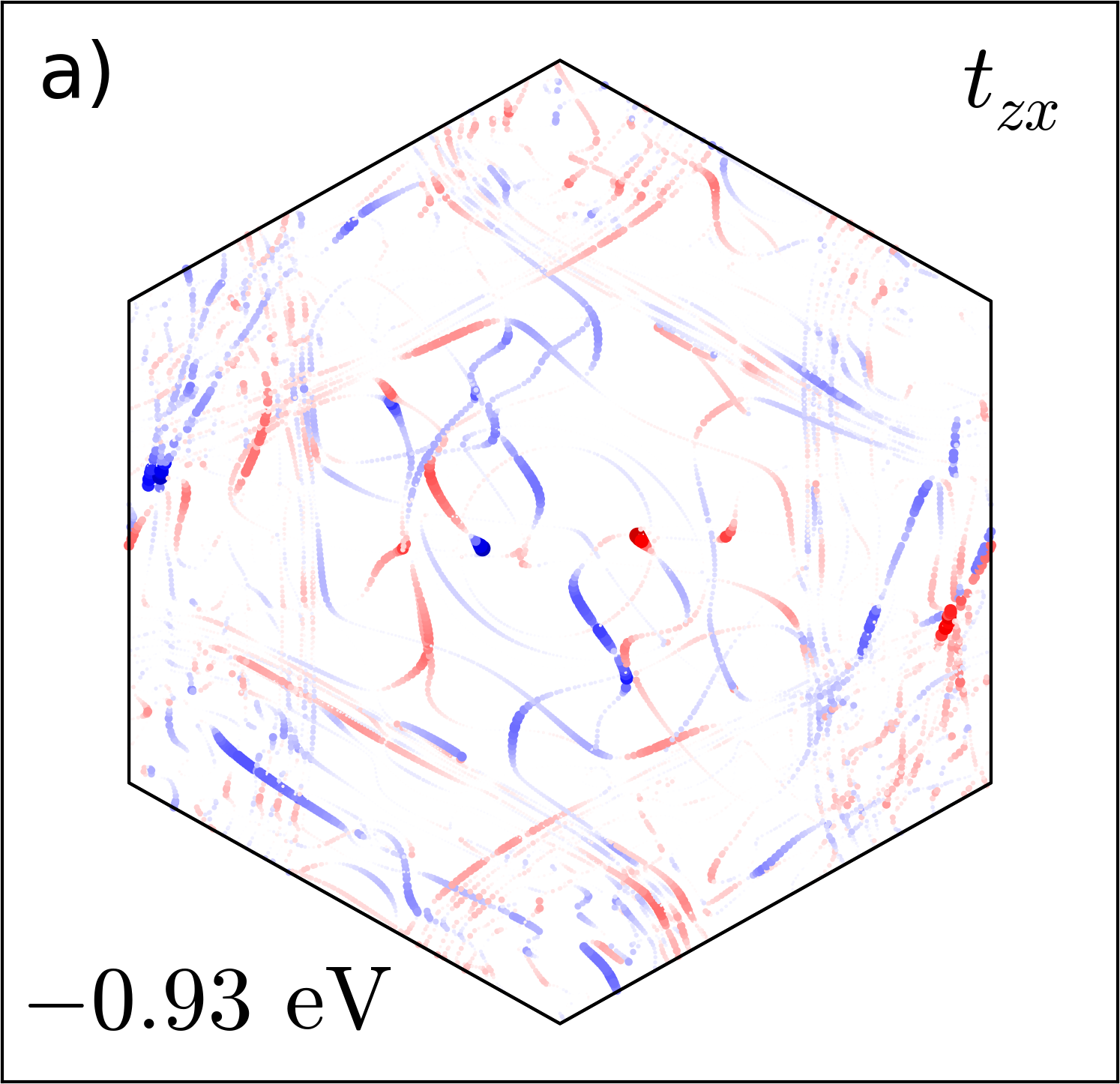}
\includegraphics*[height=3.5cm]{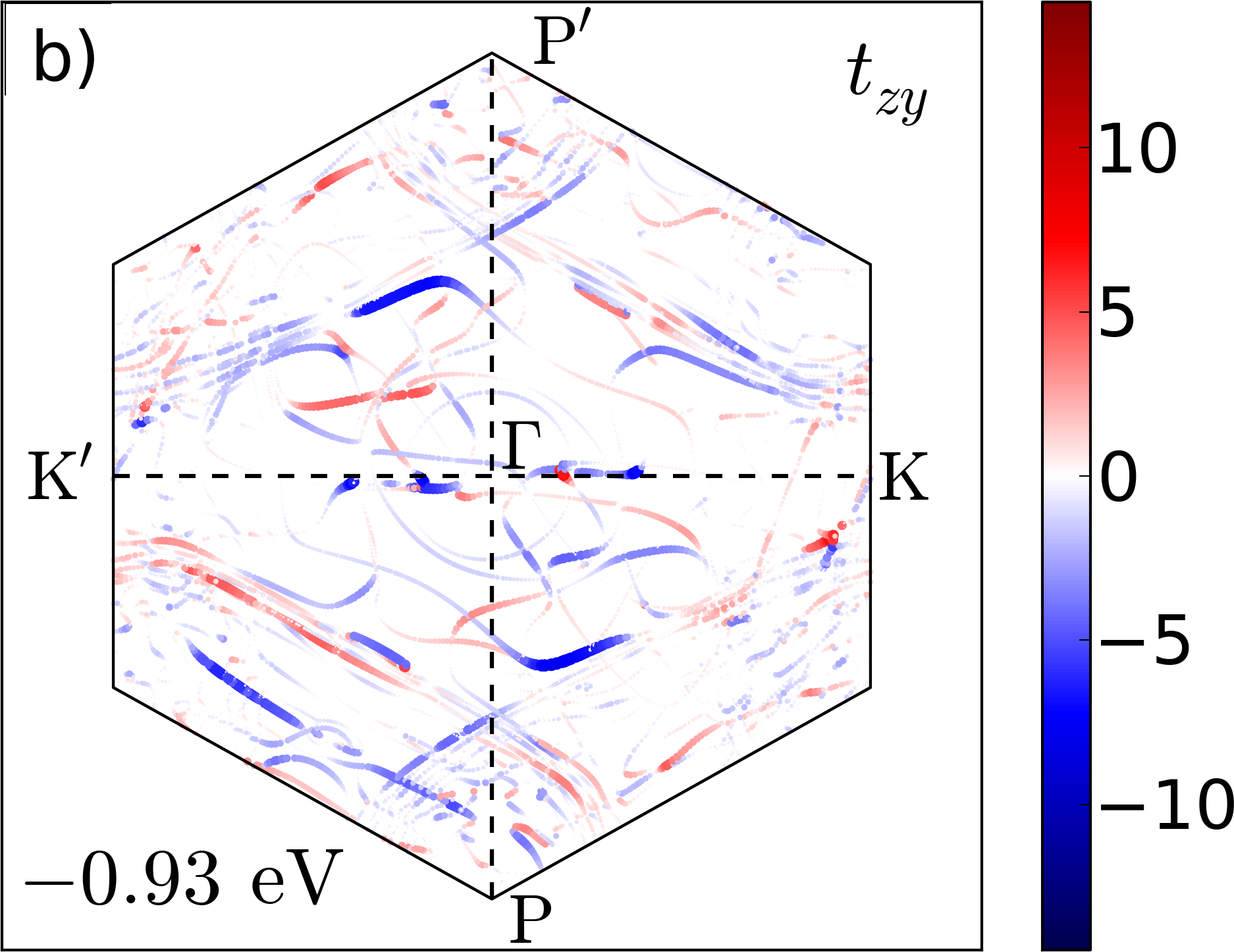}
\caption{Distribution of $\mathcal{T}_{z}(\vn{k}) \lambda_{x}(\vn{k})/|\vn{v}(\vn{k})|$ and $\mathcal{T}_{z}(\vn{k}) \lambda_{y}(\vn{k})/|\vn{v}(\vn{k})|$ in the Brillouin zone for the Bi/Ag/Fe film. The Fermi energy is set to $E_{\rm S}=E_{\rm F}-0.93$\,eV. The impurity concentration and the disorder strength are set to $\bar{c}_{imp}=0.1$ and $\Gamma = 1$\,meV, respectively. The color scale was adapted by a factor of 11.1 to account for the larger longitudinal conductivity as compared to the case where $\bar{c}_{imp}=0$ and $\Gamma = 25$\,meV (Fig.~\ref{fig_BiAgFe_t_Ef_min})}
\label{fig_BiAgFe_t_Ef_min_imp}
\end{figure}

The effect of Bi vacancies on the torkance tensor can be interpreted in terms of suppression or promotion of contributions from different states in the BZ. As a reference, we show in Fig.~\ref{fig_BiAgFe_t_Ef_min} the BZ distribution of the torkance for the case of $\Gamma=25$\,meV and $\bar{c}_{imp}=0$ (CRTA). As discussed in Sec.~\ref{sec_CRTA}, the states with large weights on the Ag$_{2}$Bi layer, which hybridize with the Rashba states, provide a very large contribution to the torkance tensor. These states are easily identified in Fig.~\ref{fig_BiAgFe_t_Ef_min} due to their Rashba-like symmetry and higher intensity as compared to other states in the BZ. For comparison, we show in Fig.~\ref{fig_BiAgFe_t_Ef_min_imp} the BZ distribution of the torkance for the case of $\Gamma=1$\,meV and $\bar{c}_{\rm imp}=0.1$. We observe a strong modification in the relative magnitude of the contributions from different regions of the BZ. In particular, the very large contributions from the states with large weights on the Ag$_{2}$Bi layer are strongly suppressed. Therefore, the large variation of the effective magnetic field observed in Fig.~\ref{fig_BiAgFe_Beff_vs_gamma}b between the cases with $\bar{c}_{imp}=0$ (CRTA) and $\bar{c}_{\rm imp}=0.1$ is driven by the suppression or promotion of the contributions from different states in the BZ. This suggests a very powerful strategy to tune the SOT in Bi/Ag/Fe thin films by proper engineering of the sources of scattering in the system.

\section{Conclusions}\label{Conclusions}
Using the KKR method, we have computed the field-like SOT in a Ag$_2$Bi-terminated Ag(111) film grown on ferromagnetic Fe(110). We demonstrated that a large part of the computed SOT arises from the SOI in the Ag$_{2}$Bi layer and therefore has a distinct non-local character, which does not originate in the intrinsic spin Hall effect. Based on the first principles transition rates induced by Bi vacancies in the Ag$_{2}$Bi layer, we compute the SOT for various types of disorder. We show that the introduction of Bi vacancies can be used as a tool to engineer the desired properties of the SOT in this system.

\section{Acknowledgements}
We gratefully acknowledge computing time on the supercomputers of J\"ulich Supercomputing Center, as well as the funding under SPP 1538 of the Deutsche Forschungsgemeinschaft. We thank Philipp R\"u{\ss}mann for discussions.

\bibliography{main}

\end{document}